\documentclass[aip,jcp,reprint,showpacs]{revtex4-1}

\usepackage{dcolumn}
\usepackage[normalem]{ulem}
\usepackage{placeins}
\usepackage{epsf}
\usepackage{subfigure}
\usepackage{epsfig,amsopn}
\usepackage{amssymb,graphicx}
\usepackage{amsmath,amsfonts}
\usepackage{graphicx}
\usepackage{bm}
\usepackage{ifpdf}
\usepackage{natbib}
\usepackage{wasysym}
\usepackage[byname]{smartref}
\usepackage{color}
\usepackage{lipsum}
\usepackage{nicefrac}
\usepackage{multirow}
\usepackage{braket}
\usepackage{float}
\usepackage{comment}
\usepackage{enumerate}
\usepackage{tabularx}
\usepackage[table]{xcolor}
\usepackage{xfrac}
\usepackage{mathtools}

\usepackage{amsmath,amsthm,verbatim,amssymb,amsfonts,amscd, graphicx}
\usepackage{braket}
\usepackage[breaklinks, colorlinks, linkcolor=blue, citecolor=blue, urlcolor=blue]{hyperref} 

\renewcommand{\v}[1]{\boldsymbol{#1}} % for vectors
\newcommand{\m}[1]{\boldsymbol{#1}} % for matrices
\newcommand{\beq}{\begin{equation}}
\newcommand{\eeq}{\end{equation}}

\newcommand{\bea}{\begin{eqnarray}}
\newcommand{\eea}{\end{eqnarray}}

\DeclareMathOperator{\cay}{cay}
\DeclareMathOperator{\Exp}{exp}
\DeclareMathOperator{\sinc}{sinc}

\begin{document}
	\title{Dimension-free path-integral molecular dynamics without preconditioning}

\author{Roman Korol}
	\affiliation{Division of Chemistry and Chemical Engineering, California Institute of Technology, Pasadena, CA 91125, USA}
\author{Jorge L. Rosa-Ra\'ices}
	\affiliation{Division of Chemistry and Chemical Engineering, California Institute of Technology, Pasadena, CA 91125, USA}
\author{Nawaf Bou-Rabee}
	\email{nawaf.bourabee@rutgers.edu}
	\affiliation{Department of Mathematical Sciences Rutgers University Camden, Camden, NJ 08102 USA}
\author{Thomas F. Miller III}
	\email{tfm@caltech.edu}
\affiliation{Division of Chemistry and Chemical Engineering, California Institute of Technology, Pasadena, CA 91125, USA}
\date{\today}

\begin{abstract}
Convergence with respect to imaginary-time discretization (i.e., the number of ring-polymer beads) is an essential part of any path-integral-based molecular dynamics (MD) calculation. 
However, an unfortunate property of existing non-preconditioned numerical integration schemes for path-integral molecular dynamics (PIMD) -- including essentially all existing ring-polymer molecular dynamics (RPMD) and thermostatted RPMD (T-RPMD) methods -- is that for a given MD timestep, the overlap between the exact ring-polymer Boltzmann distribution and that sampled using MD 
becomes zero in the infinite-bead limit.
This has clear implications for hybrid Metropolis Monte-Carlo/MD sampling schemes,
and it also causes the divergence with bead number of the primitive path-integral kinetic-energy expectation value when using standard RPMD or T-RPMD. 
We show that these and other problems can be avoided through the introduction of ``dimension-free" numerical integration schemes for which the sampled ring-polymer position distribution has non-zero overlap with the exact distribution in the infinite-bead limit for the case of a harmonic potential.  
Most notably, we introduce the  BCOCB integration scheme, which achieves dimension freedom via a particular symmetric splitting of the integration timestep and a novel implementation of the Cayley modification [J. Chem. Phys. 151, 124103 (2019)] for the free ring-polymer half-steps.
More generally, we show that dimension freedom can be achieved via  
mollification of the forces from the external physical potential. 
The  dimension-free path-integral numerical integration schemes introduced here yield finite error bounds for a given MD timestep, even as the number of beads is taken to infinity; these conclusions are proven for the case of a harmonic potential and borne out numerically for anharmonic systems  that include liquid water.
The numerical results for  BCOCB  are particularly striking, allowing for nearly three-fold increases in the stable timestep for liquid water with respect to the Bussi-Parrinello (OBABO) and Leimkuhler (BAOAB) integrators while introducing negligible errors in the calculated statistical properties and absorption spectrum. 
Importantly, the dimension-free, non-preconditioned integration schemes introduced here 
preserve
ergodicity and global second-order accuracy; and
they remain simple, black-box methods that avoid additional computational costs, tunable parameters, or system-specific implementations.

\end{abstract}
	\maketitle

%%%%%%% INTRO %%%%%%%%%

\section{Introduction}
\label{intro}

Considerable effort has been dedicated to the development of numerical integration schemes for imaginary-time path-integral molecular dynamics (PIMD).\cite{Parrinello1984} 
In comparison to standard classical molecular dynamics, PIMD numerical integration faces the additional challenge of the highly oscillatory dynamics of the  ring-polymer internal modes.
Work on PIMD numerical integration generally falls into two distinct categories.
In the first, the PIMD equations of motion are {\em preconditioned} by modifying the ring-polymer mass
matrix;\cite{Martyna1999, Minary2003, BeRoStVo2008, BePiSaSt2011, Lu2018,Zhang2017, Liu2016, BoSaActaN2018, BoEb2019}  this approach, which includes the widely used staging algorithms,\cite{Tuckerman1993} causes the integrated trajectories to differ from those of the ring-polymer molecular dynamics (RPMD) model for real-time dynamics,\cite{Craig2004,Habershon2013}
but it can lead to efficient \cite{BeRoStVo2008, BePiSaSt2011, Lu2018} 
sampling of the quantum Boltzmann-Gibbs distribution.\cite{Feynman1965,ChandlerWolynes}  In the second category, no modification is made to the ring-polymer mass matrix, i.e., the equations of motion are {\em non-preconditioned}.\cite{Habershon2013, Miller2005,Miller2005a,Ceriotti2010,Ceriotti2011,Rossi2014,Rossi2018}

With the aim of providing useful models for real-time quantum dynamics, as well as simple and efficient algorithms for equilibrium thermal sampling, the current work focuses on  non-preconditioned PIMD numerical integration, notable examples of which include RPMD\cite{Craig2004,Habershon2013} 
and its thermostatted variant T-RPMD.\cite{Rossi2014}  
Numerical integration schemes for the latter methods typically employ symmetric factorizations of the time-evolution operator of the form\cite{Tuckerman1993,Miller2005,Miller2005a,Bussi2007,BoOw2010,BoVa2014,Ceriotti2010,Ceriotti2011,Rossi2014,Rossi2018,Leimkuhler2013} 
\begin{equation}
\begin{aligned}
e^{\Delta t \mathcal{L}} \approx
 e^{a \frac{\Delta t}{2}\mathcal{O}}
e^{\frac{\Delta t}{2}\mathcal{B}}
e^{\frac{\Delta t}{2} \mathcal{A}}
e^{(1-a) \Delta t\mathcal{O}}  
e^{\frac{\Delta t}{2} \mathcal{A}}
e^{\frac{\Delta t}{2}\mathcal{B}}
 e^{a \frac{ \Delta t}{2}\mathcal{O}}
 \end{aligned}
	\label{eq:splitwithnoise}
\end{equation} 
\begin{comment}
\begin{equation}
\begin{drcases}
 e^{\frac{\Delta t}{2}\mathcal{O}}
e^{\frac{\Delta t}{2}\mathcal{B}}
e^{\Delta t \mathcal{A}}
e^{\frac{\Delta t}{2}\mathcal{B}}
e^{\frac{\Delta t}{2}\mathcal{O}}  \\
e^{\frac{\Delta t}{2}\mathcal{B}}
e ^{\frac{\Delta t}{2}\mathcal{A}}
e^{\Delta t \mathcal{O}}
e^{\frac{\Delta t}{2}\mathcal{A}}
e^{\frac{\Delta t}{2}\mathcal{B}} 
\end{drcases}
= e^{\Delta t \mathcal{L}} + O(\Delta t^3)
	\label{eq:splitwithnoise}
\end{equation} 
\end{comment}
where the operator $\mathcal{L}=\mathcal{A}+\mathcal{B}+\mathcal{O}$
includes contributions from the purely harmonic free ring-polymer motion $\mathcal{A}$, the external  potential $\mathcal{B}$, and a thermostat $\mathcal{O}$. Note that the standard microcanonical RPMD numerical integration scheme is recovered in the limit of zero coupling to the thermostat, and that Eq.~\ref{eq:splitwithnoise} yields the    ``OBABO'' scheme of Bussi and Parrinello\cite{Bussi2007} when $a=1$ and the ``BAOAB'' scheme of Leimkuhler\cite{Leimkuhler2013} when $a=0$.

In our previous work,\cite{KoBoMi2019} we emphasized that earlier PIMD numerical integration schemes had overlooked a fundamental aspect of the $\exp((\Delta t/2) \mathcal{A})$ sub-step of the time evolution in Eq.~\ref{eq:splitwithnoise}.
Standard practice in these integration schemes has been to exactly evolve the harmonic free ring-polymer dynamics associated with $\exp((\Delta t/2) \mathcal{A})$  using the uncoupled free ring-polymer normal modes,\cite{Tuckerman1993,Miller2005,Miller2005a,Ceriotti2010}  
which was shown to lack the property of strong stability in the numerical integration, leading to resonance instabilities for  microcanonical RPMD and loss of ergodicity for T-RPMD.\cite{KoBoMi2019}  Use of the Cayley modification to the free ring-polymer motion was shown to impart strong stability to the time evolution,
thereby 
improving numerical stability for  microcanonical RPMD and restoring ergodicity for T-RPMD.\cite{KoBoMi2019}

In the current study, we focus on the accuracy of both statistical and dynamical properties of the OBABO and BAOAB schemes, as well as the corresponding integrators obtained when the exact free ring-polymer step is replaced by the strongly stable Cayley modification (OBCBO and BCOCB, respectively).  Particular attention is paid to the effect of finite-timestep error with these integrators in the limit of large bead numbers.
Of these four integrators, it is found that only BCOCB is ``dimension-free,'' in the sense that the sampled ring-polymer position distribution has non-zero overlap with the exact distribution in the infinite-bead limit for the case of a harmonic potential.  
It is further shown that the OBCBO scheme can be made dimension-free via the technique of force mollification. 
It is shown that the newly introduced BCOCB integrator yields  better accuracy than all other considered non-preconditioned PIMD integrators and allows for substantially larger timesteps in the calculation of both statistical and dynamical properties. Importantly, these gains are made without loss of computational efficiency or  algorithmic simplicity.

\section{Non-preconditioned PIMD} 
\label{motivation}

Consider a one-dimensional molecular system with potential energy function $V(q)$ and mass $m$.  The equations of motion for the corresponding $n$-bead ring polymer held at constant temperature $T$ by a Langevin thermostat are
\begin{equation} \label{eq:trpmd}
\begin{aligned}
    \dot{\v{q}}(t) = \v{v}(t) \;, ~
    &\dot{\v{v}}(t) = - \m{\Omega}^2 \v{q}(t) + \frac{1}{m_n} \v{F}( \v{q}(t) ) \\
    & \quad - \m{\Gamma} \v{v}(t) + \sqrt{\frac{2}{\beta m_n}} \m{\Gamma}^{1/2} \dot{\v{W}}(t) \;.
\end{aligned}
\end{equation}  
Here,  $\v{W}$ is an $n$-dimensional standard Brownian motion; $\v{q}(t) = (q_0(t), \dots, q_{n-1}(t))$ is the vector of positions for the $n$ ring-polymer beads
at time $t \ge 0$ and $\v{v}(t)$ are the corresponding velocities; $m_n = m/n$ and $\beta = (k_B T)^{-1}$; and $\v{F}( \v{q}) =  -\nabla V^{\textrm{ext}}_n( \v{q})$, where $V^{\textrm{ext}}_n$ is the contribution of the external  potential, 
\beq
V^{\textrm{ext}}_n(\v{q}) = \frac{1}{n} \sum_{j=0}^{n-1} V(q_j) \;.
	\label{eq:Vext}
\eeq
Moreover, $\m{\Omega}^2$ is the following $n \times n$ symmetric positive semi-definite matrix 
 \beq
	 \m{\Omega}^2 = -\kappa_n^2 \begin{bmatrix} 
	-2 & 1 & 0 & \cdots & 0 & 1 \\
1 & -2 & 1 & 0 & \cdots & 0 \\
 & \ddots & \ddots & \ddots \\
  && \ddots & \ddots & \ddots \\
0 & \cdots & 0 & 1 & -2 & 1 \\
1 & 0 & \cdots & 0 & 1 & -2
	\end{bmatrix}  \;,  
	\label{eq:L} 
\eeq
where $\kappa_n =  n/(\hbar\beta)$.  Note that  $\m{\Omega}$ can be diagonalized by an $n \times n$ orthonormal real discrete Fourier transform matrix $\m{U}$ as follows
\beq \label{eq:VDV}
\m{\Omega} =  \m{U} \operatorname{diag}(0, \omega_{1,n},  \dots, \omega_{n-1,n}) \m{U}^{\mathrm{T}}, 
\eeq 
where $\omega_{j,n}$ is the $j$th Matsubara frequency\cite{Matsubara1955}
given by \begin{equation} \label{eq:eigenvalues}
\omega_{j,n} = \begin{cases}
2 \kappa_n \sin\left( \frac{\pi j}{2 n} \right)
 & \text{if $j$ is even} \;, \\
2 \kappa_n \sin\left( \frac{\pi (j+1)}{2 n} \right)  & \text{else} \;.
\end{cases}
\end{equation}
Finally, the matrix $\m{\Gamma}$ in Eq.~\ref{eq:trpmd} is typically an $n \times n$ symmetric positive semi-definite friction matrix of the form
\beq 
\m{\Gamma} =  \m{U} \operatorname{diag}(0, \gamma_{1},  \dots, \gamma_{n-1}) \m{U}^{\mathrm{T}}, 
\eeq
where $\gamma_j$ is the friction factor in the $j$th normal mode.

In RPMD and T-RPMD calculations, one is often interested in the dynamics of Eq.~\ref{eq:trpmd} with initial conditions drawn from the stationary distribution with non-normalized density 
$\exp(- \beta H_n(\v{q},\v{v}) )$, where $H_n(\v{q}, \v{v})$ is the ring-polymer Hamiltonian defined by \beq
	H_n(\v{q},\v{v})=H_n^0(\v{q},\v{v})+
V^{\textrm{ext}}_n(\v{q}),
	\label{eq:H_n}
\eeq  and $H_n^0(\v{q}, \v{v}) = (1/2) m_n \left(  |\v{v}|^2 +\v{q}^{\mathrm{T}} \m{\Omega}^2 \v{q} \right)$ is the free ring-polymer Hamiltonian.

The standard method for discretizing Eq.~\ref{eq:trpmd} is to use a symmetric splitting method of the form of Eq.~\ref{eq:splitwithnoise} that consists of a combination of three types of sub-steps:
(i) exact free ring-polymer evolution of timestep $\tau$, \begin{equation} \label{eq:L0}
\begin{pmatrix} \v{q} \\ \v{v} \end{pmatrix} \leftarrow \exp(\tau \m{A})  \begin{pmatrix} \v{q} \\ \v{v} \end{pmatrix}, \end{equation}
where $\m{A} = \begin{bmatrix} \m{0}  & \m{I} \\ -\m{\Omega}^2 & \m{0} \end{bmatrix} $ is the Hamiltonian matrix associated to the free ring polymer, 
(ii) velocity updates of timestep $\tau$ due to forces from the external potential, \begin{equation} \label{Bsubstep}  \v{v} \leftarrow \v{v} + \tau \frac{1}{m_n} \v{F}(\v{q}),
\end{equation} 
and (iii) velocity updates of timestep $\tau$ due to the thermostat, 
\begin{equation} \label{Osubstep}  \v{v} \leftarrow \exp(-\tau \m{\Gamma} ) \v{v} + \sqrt{\frac{1}{\beta m_n}} (\m{I} - \exp(-2\tau \m{\Gamma}) )^{1/2} \v{\xi},
\end{equation}
where $\m{I}$ is the $n \times n$ identity matrix and $\v{\xi}$ is an $n$-dimensional vector whose components are independent, standard normal random variables. 
The acronyms OBABO and BAOAB indicate the order in which these sub-steps are applied, as indicated in Eq.~\ref{eq:splitwithnoise} with $a=1$ or $a=0$, respectively.

In previous work,\cite{KoBoMi2019} we showed that the matrix exponential for the free ring-polymer evolution in Eq.~\ref{eq:L0} is not a strongly stable symplectic matrix, and as a consequence, the 
the OBABO and BAOAB schemes 
can display non-ergodicity at timesteps $\Delta t = k \pi / \omega_{j,n}$ for any $1 \le j \le n$ and $k \ge 1$.  We also identified a maximum safe timestep size $\Delta t_{\star} = \beta \hbar \pi / (2 n)$, below which the matrix exponential is strongly stable.  As $n \to \infty$, this maximum safe timestep goes to zero, such that no finite timestep for the scheme in Eq.~\ref{eq:splitwithnoise} is safe in this limit from non-ergodicity.

This non-ergodicity motivates the Cayley modification\cite{KoBoMi2019} which consists of approximating the matrix exponential appearing in Eq.~\ref{eq:L0} with the Cayley transform.  Specifically, for the Cayley-modified OBABO scheme (called OBCBO), we replace the exact free-ring polymer update of timestep $\tau=\Delta t$  with
\begin{equation} 
\cay(\Delta t \m{A}) = (\m{I} - (1/2) \Delta t \m{A})^{-1} (\m{I} + (1/2) \Delta t \m{A}) \;. \label{eq:cay} 
\end{equation}
For the Cayley-modified BAOAB scheme (called BCOCB), we replace the two exact free ring-polymer updates of half-timestep $\tau=\Delta t/2$  with $\cay(\Delta t \m{A})^{1/2}$.  While it might be expected that these half-timestep updates would instead be replaced with $\cay((\Delta t/2) \m{A})$, such a choice leads to a loss of strong stability.  Our use of the square root of the Cayley transform 
preserves strong stability, symplecticity, time reversibility, local third-order accuracy, and by definition satisfies $\cay(\Delta t \m{A})^{1/2} \cay(\Delta t \m{A})^{1/2} = \cay(\Delta t \m{A})$.  Furthermore, the square root of the Cayley transform is  no more complicated to evaluate than the Cayley transform itself.  Both the OBCBO and BCOCB Cayley modifications of  Eq.~\ref{eq:splitwithnoise} are ergodic for a fixed timestep, irrespective of the number of beads; moreover, like Eq.~\ref{eq:splitwithnoise}, the Cayley modified integrators exhibit locally third-order accuracy in the timestep and leave invariant the free ring-polymer Boltzmann-Gibbs distribution in the special case of a constant external potential ($V\equiv \textrm{const.}$).\cite{KoBoMi2019}

\section{BCOCB Avoids Pathologies in the Infinite Bead limit}
\label{sec:BCOCB}

In this section, we show that of the OBABO, BAOAB, OBCBO, and BCOCB integration schemes, only  BCOCB  is dimension-free. 
Although the current section presents analytical results for the specific case of a harmonic external potential, these results are  supported by numerical results for anharmonic external potentials in the subsequent sections.

To this end, consider the $j^{\textrm{th}}$
internal ring-polymer mode with frequency $\omega_{j,n}$, in the presence of a harmonic external potential $V(q) =  (1/2) \Lambda q^2$ and a Langevin thermostat with friction $\gamma_j$. 
Expressed in terms of the normal mode coordinates, obtained from the Cartesian positions and velocities via the orthogonal transformation
	\beq \label{eq:nm_transform}
	\v{\varrho} =  \m{U}^{\mathrm{T}} \v{q} \qquad \text{and} \qquad 
    	\v{\varphi} =  \m{U}^{\mathrm{T}} \v{v}
	\eeq
	where $\m{U}$ is defined in Eq.~\ref{eq:VDV}, the non-preconditioned PIMD equations of motion for this mode are
\begin{equation} \label{eq:langevin_equations_1D}
\begin{aligned}
\begin{bmatrix} \dot \varrho_j(t) \\ \dot \varphi_j(t) \end{bmatrix} &=\m{K}_j \begin{bmatrix} \varrho_j(t) \\ \varphi_j(t) \end{bmatrix} + \begin{bmatrix} 0 \\ \sqrt{2\beta^{-1} m_n^{-1} \gamma_j} \dot W_j(t) \end{bmatrix}    \\
 \m{K}_j &= \m{A}_j + \m{B} + \m{O}_j,
 \end{aligned}
\end{equation}  
where $\dot W_j$ is a scalar white-noise, and we have introduced the following $2 \times 2$ matrices \[
\m{A}_j = \begin{bmatrix} 0  &1\\ -\omega_{j,n}^2  & 0 \end{bmatrix} , ~ 
\m{B} = \begin{bmatrix} 0  & 0 \\ -\Lambda / m & 0 \end{bmatrix} , ~ \text{and} ~
\m{O}_j = \begin{bmatrix} 0  & 0 \\ 0 & -\gamma_j \end{bmatrix} .
\]  
The solution $(\varrho_j(t), \varphi_j(t))$ of Eq.~\ref{eq:langevin_equations_1D} is a bivariate Gaussian, 
and in the limit as $t \to \infty$, the probability distribution of $(\varrho_j(t), \varphi_j(t))$ converges  to a centered bivariate normal distribution with covariance matrix  
\begin{equation} \label{eq:exact_IM_1D}
   \m{\Sigma}_j = \frac{1}{\beta m_n} \begin{bmatrix} s_j^2 & 0 \\ 0 & 1 \end{bmatrix} \;, \qquad  s_j^2 = \frac{1}{\Lambda/m + \omega_{j,n}^{2}} \;.
\end{equation}

For this system, a single timestep of Eq.~\ref{eq:splitwithnoise} can be compactly written as  \begin{equation} \label{eq:trpmd_schemes_1D} \begin{bmatrix} \varrho_j(t+\Delta t) \\ \varphi_j(t+\Delta t) \end{bmatrix}  =  \m{M}_j \begin{bmatrix} \varrho_j(t) \\ \varphi_j(t) \end{bmatrix} +  \m{R}_j^{1/2} \begin{bmatrix} \xi_0 \\ \eta_0 \end{bmatrix},
\end{equation}
where $\xi_0$ and $\eta_0$ are independent standard normal random variables, and we have introduced the following $2 \times 2$ matrices
\begin{align*}
&\m{M}_j = 
e^{ a \frac{\Delta t}{2} \m{O}_j} e^{ \frac{\Delta t}{2} \m{B}} e^{\frac{\Delta t}{2} \m{A}_j}  e^{ (1-a) \Delta t \m{O}_j} 
 e^{ \frac{\Delta t}{2} \m{A}_j} 
 e^{ \frac{\Delta t}{2} \m{B}} e^{ a \frac{\Delta t}{2} \m{O}_j}
\\
&\m{R}_j = \frac{1-e^{-2 (1-a) \gamma_j \Delta t} }{\beta m_n}  \m{N}_j \m{P} \m{N}_j^{\mathrm{T}} \\ 
& ~~ + \frac{1-e^{-a \gamma_j \Delta t} }{\beta m_n}    \left(  ( \m{M}_j e^{ -a \frac{\Delta t}{2} \m{O}_j}) \m{P} (\m{M}_j e^{ -a \frac{\Delta t}{2} \m{O}_j} )^{\mathrm{T}} + \m{P} \right) 
\end{align*}
where
$\m{P} = \begin{bmatrix} 0 & 0 \\ 0 & 1 \end{bmatrix}$ and $\m{N}_j = e^{ a \frac{\Delta t}{2} \m{O}_j} e^{ \frac{\Delta t}{2} \m{B}} e^{\frac{\Delta t}{2} \m{A}_j}$.
The corresponding step for the Cayley modification is obtained by replacing $\Exp( (\Delta t/2) \m{A}_j)$ in Eq.~\ref{eq:trpmd_schemes_1D} with  $\cay(\Delta t \m{A}_j)^{1/2}$, which is given by \begin{equation}
\cay(\Delta t \m{A}_j)^{1/2} = \sqrt{\frac{1}{4 + \omega_{j,n}^2 \Delta t^2}}  \begin{bmatrix} 2 &  \Delta t \\ -\omega_{j,n}^2 \Delta t & 2 \end{bmatrix}   \;.
\end{equation}
A sufficient condition\footnote{In the special case when $\Lambda=0$, the given condition for OBCBO corrects a sign error in Eq.~37 of Ref.~\cite{KoBoMi2019}.} for ergodicity of Eq.~\ref{eq:trpmd_schemes_1D} is 
\begin{equation} \label{eq:stab_1d}
\quad\qquad 1 > \mathsf{A}_{j,\Delta t}^2 \cosh^2((\Delta t / 2) \gamma_j ) \;,
\end{equation} 
where 
\[
    \mathsf{A}_{j,\Delta t} = 
 \cos(\Delta t \omega_{j,n}) - \dfrac{(\Lambda/m) \Delta t}{2 \omega_{j,n}} \sin(\Delta t \omega_{j,n})  \;.
    \]
For the Cayley modification of Eq.~\ref{eq:trpmd_schemes_1D}, Eq.~\ref{eq:stab_1d} still provides a sufficient condition for ergodicity, except with
\[ \mathsf{A}_{j,\Delta t} = 
-1 + \dfrac{8 - 2 (\Lambda/m) \Delta t^2}{4 + \omega_{j,n}^2 \Delta t^2} \;.
\]
Due to the lack of strong stability in the exact free ring-polymer evolution, Eq.~\ref{eq:trpmd_schemes_1D} fails to meet the condition in Eq.~\ref{eq:stab_1d} and becomes non-ergodic whenever $\Delta t = k \pi / \omega_{j,n}$ where $k \ge 1$;\cite{KoBoMi2019} no such problem exists for the Cayley modification.
Regardless, assuming that the condition in Eq.~\ref{eq:stab_1d} holds, the numerical stationary distribution is a centered Gaussian with $2 \times 2$ covariance matrix $\m{\Sigma}_{j,\Delta t}$ that satisfies the linear equation \begin{align*}
&\quad \qquad \m{\Sigma}_{j,\Delta t} = \m{M}_j  \m{\Sigma}_{j,\Delta t} \m{M}_j^{\mathrm{T}} + \m{R}_j \;,
\end{align*} for which the solution is 
\begin{equation} \label{eq:numerical_IM_1D_a}
\m{\Sigma}_{j,\Delta t} = \frac{1}{\beta m_n} \begin{bmatrix} s_{j, \Delta t}^2  & 0 \\ 0 & r_{j,\Delta t}^2 \end{bmatrix}
\end{equation}
where the variance in the position and velocity marginal are $(\beta m_n)^{-1} s_{j, \Delta t}^2$ and $(\beta m_n)^{-1} r_{j, \Delta t}^2$  with \begin{align} \label{eq:numerical_IM_1D_OBAOABO_s}
s_{j, \Delta t}^2 &= \begin{cases}
\dfrac{1}{\omega_{j,n}^2 +  \frac{\Lambda \Delta t \omega_{j,n}}{m} \cot(\Delta t \omega_{j,n})  -(\frac{\Lambda \Delta t}{2 m})^2}  & a =1 \\
\dfrac{1}{\omega_{j,n}^2 +  \frac{\Lambda \Delta t \omega_{j,n}}{2 m} \cot(\frac{\Delta t}{2} \omega_{j,n})}   & a=0 
\end{cases} \\
\label{eq:numerical_IM_1D_OBAOABO_r}
r_{j, \Delta t}^2 &= \begin{cases}
1 & a=1 \\
\dfrac{2 m \omega_{j,n} - \Lambda \Delta t \tan(\frac{\Delta t}{2} \omega_{j,n}) }{2 m \omega_{j,n}}   & a=0 
\end{cases}
\end{align}
For the Cayley modification of Eq.~\ref{eq:trpmd_schemes_1D},
\begin{align}
\label{eq:numerical_IM_1D_OBCOCBO_s}
s_{j, \Delta t}^2 &= 
\dfrac{4 m}{4 m - a \Delta t^2 \Lambda} s_j^2  \;, \\
\label{eq:numerical_IM_1D_OBCOCBO_r}
r_{j, \Delta t}^2 &= 
\dfrac{4 m - (1-a) \Delta t^2 \Lambda}{4 m}   \;.
\end{align}
Note that these numerical stationary distributions are independent of
the friction parameter $\gamma_j$, which is a  benefit of schemes based on splitting the T-RPMD dynamics into Hamiltonian and thermostat parts, and using the exact Ornstein-Uhlenbeck flow in Eq.~\ref{Osubstep} to evolve the thermostat part.  Moreover,  comparing the exact covariance matrix in Eq.~\ref{eq:exact_IM_1D} with the finite-timestep approximations in Eqs.~\ref{eq:numerical_IM_1D_a}-\ref{eq:numerical_IM_1D_OBCOCBO_r}, note that in all cases $\m{\Sigma}_{j} = \lim_{\Delta t \to 0} \m{\Sigma}_{j,\Delta t}$.
These results have previously been reported for the OBABO (Eqs.~\ref{eq:numerical_IM_1D_OBAOABO_s} and \ref{eq:numerical_IM_1D_OBAOABO_r}, $a=1$) and BAOAB (Eqs.~\ref{eq:numerical_IM_1D_OBAOABO_s} and \ref{eq:numerical_IM_1D_OBAOABO_r}, $a=0$) schemes\cite{Li2017, Liu2016} but not for the OBCBO (Eqs.~\ref{eq:numerical_IM_1D_OBCOCBO_s} and \ref{eq:numerical_IM_1D_OBCOCBO_r}, $a=1$) or BCOCB (Eqs.~\ref{eq:numerical_IM_1D_OBCOCBO_s} and \ref{eq:numerical_IM_1D_OBCOCBO_r}, $a=0$) schemes. 

In the infinite bead limit, the exact and numerical position-marginals can be written as an infinite product of one-dimensional centered normal distributions with variances given by $(\beta m_n)^{-1} s_{j}^2$  and $(\beta m_n)^{-1} s_{j, \Delta t}^2$, respectively.  By Kakutani's theorem, \cite{kakutani1986equivalence,bogachev1998gaussian} these two distributions have a non-zero overlap if and only if the following series converges, \begin{equation} \label{eq:kakutani_series}
\sum_{j=1}^{\infty} 
\left( 1-\frac{s_{j}}{s_{j,\Delta t}}  \right)^2  \;.
\end{equation}
For OBABO and BAOAB, due to the oscillatory cotangent term appearing in $s_{j,\Delta t}$, the limit $\lim_{j \to \infty} ( 1-s_{j}/s_{j,\Delta t} )  ^2$ does not exist, and therefore, the series does not converge. 
For OBCBO, the $j$th summand of this series is \[
\frac{\Delta t^4 \Lambda^2}{16 m^2} \left(1+\sqrt{ \frac{4 m - \Delta t^2 \Lambda}{4 m}}\right)^{-2} \;,
\] which more obviously leads to a divergent series. Therefore, for  OBABO, OBCBO, and BAOAB, the numerical stationary distribution has no overlap with the exact stationary distribution in the infinite bead limit; it is in this sense that these schemes fail to exhibit the property of dimensionality freedom. 
Remarkably, BCOCB is exact in the position marginal and thus exhibits dimensionality freedom.
See Appendix~A for a brief summary of the properties of other symmetric splittings that were considered.

\section{Consequences for the primitive kinetic energy expectation value }

In the current section, we show that the non-overlap pathology of the OBABO, BAOAB, and OBCBO schemes causes a divergence with increasing bead number of the primitive path-integral kinetic-energy expectation value,
an issue that is numerically well known for OBABO and BAOAB.\cite{Li2017, Liu2016, Perez2011, Marsalek2014} 
We further show that this divergence is fully eliminated via the BCOCB scheme -- as expected.

\begin{figure}
\begin{center}
\includegraphics[width=0.48\textwidth]{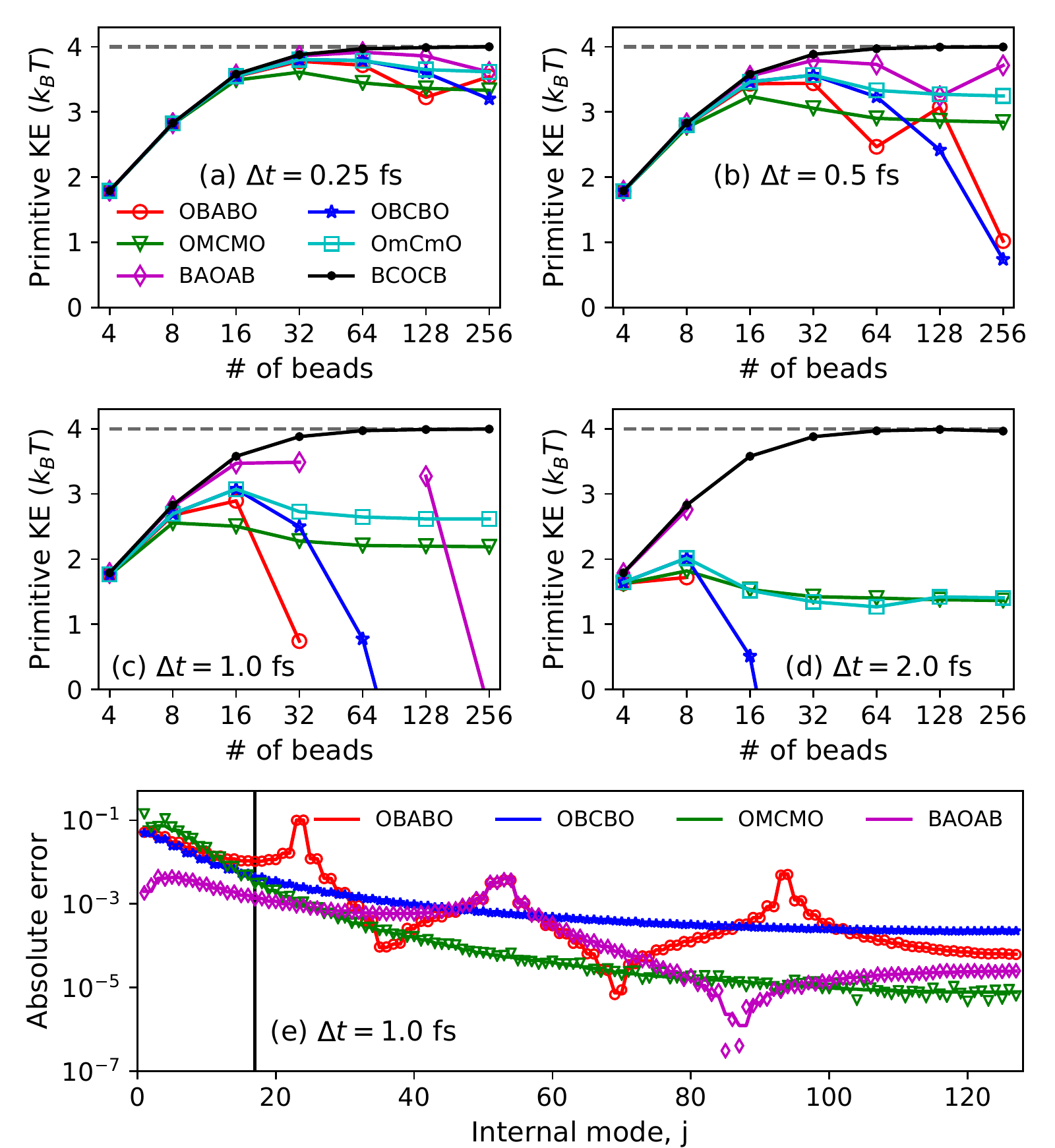}
\end{center}
\caption{\small  {\bf Primitive kinetic energy expectation values}
for a harmonic potential $V(q)=\frac12 \Lambda q^2$ with $\Lambda=256$, $\hbar=m=1$, and reciprocal temperature $\beta=1$; choosing energies to be in units of $k_\textrm{B}T$ at room temperature ($300$ K), then  $\beta\hbar\approx25.5$ fs and $\Lambda=m\omega^2$ where  $\omega=3315$ cm$^{-1}$.
 (a-d) For various MD timesteps, the primitive kinetic energy expectation value as a function of the number of ring-polymer beads, with the exact kinetic energy indicated as a dashed gray line. The standard error of all visible data points in each plot is smaller than the symbol size.
 (e) Per-mode error in the variance of position coordinate of the normal modes for simulations run with 128 ring-polymer beads and a timestep of 1 fs; solid lines are analytic predictions from Eq.~\ref{eq:permodevarerror} with \ref{eq:numerical_IM_1D_OBAOABO_s} and \ref{eq:numerical_IM_1D_OBCOCBO_s} defining $s^2_{j,\Delta t}$ for the different schemes; points indicate the results of numerical PIMD simulations using the various integration schemes. The BCOCB scheme is not shown since it has zero error for all internal modes. The black vertical line indicates the crossover frequency ($\omega_{\mathrm{x}} =  2/\Delta t$) for the error of  OBCBO and OMCMO  based on the bounds in Eqs.~\ref{j_error_OBCBO} and \ref{j_error_OMCMO}. 
}
  \label{fig:logError}
\end{figure}

The primitive kinetic energy expectation value is given by\cite{Barker1979, Herman1982}
\begin{eqnarray}
\braket{KE_{\textrm{prim}}} &=& \frac{n}{2\beta}-\sum_{j=1}^n\frac{m_n \kappa_n^2}{2}\braket{(q_j-q_{j-1})^2}\\
&=& \frac{1}{2\beta} + \sum_{j=1}^{n-1}\left(\frac{1}{2\beta} - \frac{m_n \omega_{j,n}^2}{2}\braket{\varrho_j^2}\right)
\label{eq:primitive}
\end{eqnarray}
where the first equality involves a sum over the ring-polymer beads in  Cartesian coordinates (with $q_n = q_0$), and the second equality performs the summation in terms of the ring-polymer normal modes.  
The divergence of this expectation value is numerically illustrated for the simple case of a harmonic oscillator (Figs.~\ref{fig:logError}a-d); note that for larger MD timesteps, the OBABO, BAOAB, and OBCBO schemes fail to reach a plateau with increasing bead number and dramatically deviate from the exact result (dashed line).
The same divergence for OBABO and BAOAB has been  numerically observed in many systems,\cite{Li2017, Liu2016, Perez2011,Marsalek2014} 
 including  liquid water which we discuss later. 
 A striking observation from Figs.~\ref{fig:logError}a-d is that the BCOCB exhibits no such divergence or error in the primitive kinetic energy expectation value at high bead number, regardless of the employed timestep.

Using Eq.~\ref{eq:exact_IM_1D}, note that the contribution to the primitive kinetic energy  expectation value from the $j$th ring-polymer mode is \[
\braket{KE_j}=\frac{1}{2\beta}\left(1-\omega_{j,n}^2 s_j^2 \right),
\]
such that in the infinite-bead limit,
\begin{equation} \label{eq:ke_infinite_bead}
\begin{aligned}
\lim_{n \to \infty} \sum_{j=0}^{n-1} \braket{KE_j} =  \frac{\hbar}{4} \sqrt{\frac{ \Lambda}{m}} \left( 1 + \frac{2}{e^{\hbar \beta \sqrt{\Lambda/ m} }-1} \right) \;.
\end{aligned}
\end{equation}
Similarly using Eq.~\ref{eq:numerical_IM_1D_a}, the $j$th-mode contribution to the kinetic energy from the finite-timestep numerical expectation value is  
\begin{align}
    \braket{KE_{j}}_{\Delta t}=\frac{1}{2\beta}\left(1-\omega_{j,n}^2 s_{j,\Delta t}^2 \right).
\end{align}
Thus, the per-mode error in kinetic energy is  
 \begin{align}
|  \braket{KE_j} -  \braket{KE_{j}}_{\Delta t}| =\frac{m_n \omega_{j,n}^2}{2}  \rho_{j,\Delta t}\;, 
\label{eq:permodeerror}
\end{align}
where the per-mode error in the position marginal for internal mode $j$ is
 \begin{align}
\rho_{j,\Delta t}=\frac{1}{\beta m_n} \left| s_{j}^2- s_{j,\Delta t}^2 \right| \;, 
\label{eq:permodevarerror}
\end{align}
where $s_{j,\Delta t}$ is given by Eq.~\ref{eq:numerical_IM_1D_OBAOABO_s} for the cases of 
OBABO ($a=1$) and BAOAB ($a=0$) and by  Eq.~\ref{eq:numerical_IM_1D_OBCOCBO_s} for the cases of OBCBO ($a=1$) and BCOCB ($a=0$). Note that this error vanishes only for the BCOCB scheme, which satisfies $\rho_{j,\Delta t}=0$ for each mode $j$, irrespective of the timestep $\Delta t$. 

Eqs.~\ref{eq:permodeerror} and \ref{eq:permodevarerror} indicate that the primitive kinetic energy estimator is a sensitive measure of the finite-timestep error in the sampled ring-polymer position distribution associated with the high-frequency modes.  
Fig.~\ref{fig:logError}e
resolves this per-mode error, $\rho_{j,\Delta t}$, for each internal mode in simulations that employ a total of 128 beads, including results from 
OBABO (red), BAOAB (magenta) and OBCBO (blue)  using a timestep of 1 fs, with the solid lines indicating the analytical predictions in Eq.~\ref{eq:permodevarerror}
and with the  dots indicating the result of numerical simulations.  The analytical results are  fully reproduced by the simulations.
Note that the OBABO per-mode error exhibits dramatic spikes for $\omega_{j,n}\Delta t= k\pi$ where $1 \le j \le n$ and for some $k \ge 1$,
which coincide with the loss of ergodicity of that integration scheme. The BAOAB scheme  exhibits these resonance instabilities at even values of $k$.
However, it is the failure of this per-mode error to sufficiently decay
as a function of the mode number for all three of OBABO, BAOAB and OBCBO that gives rise upon summation  to the divergence of the primitive kinetic energy expectation value, as seen for this particular timestep value in Fig.~\ref{fig:logError}d.
Since $\omega_{j,n}^2 s_{j}^2 \to 1$ as $n \to \infty$, the convergence of $\sum_{j=1}^{\infty} |  \braket{KE_j} -  \braket{KE_{j}}_{\Delta t}|$  reduces to the convergence of the series $\sum_{j=1}^{\infty} \left| s_{j}^2 - s_{j,\Delta t}^2  \right|$, which 
diverges for both OBABO and OBCBO due to the  same reasons as discussed in the previous section.

\section{Dimensionality Freedom for OBCBO via Force Mollification}

The previous sections have demonstrated that whereas the BCOCB integrator exhibits dimensionality freedom, the OBCBO integrator does not.  In the current section, we show that this shortcoming of OBCBO can be addressed by the use of force mollification, 
in which the  external potential energy in Eq.~\ref{eq:Vext} 
is replaced by 
\begin{equation} \label{eq:Vextmolly}
\tilde V^{\textrm{ext}}_n( \v{q}) = V^{\textrm{ext}}_n( \sinc(\m{\tilde \Omega } \Delta t/2 ) \v{q}),
\end{equation} 
where $ \m{\tilde \Omega}$ is any positive semi-definite $n\times n$ matrix that has the same eigenvectors as $\m{\Omega}$ (Eq.~\ref{eq:VDV}) while possibly having different eigenvalues.
Force mollification has not previously been employed for PIMD, although the strategy  
originates from a variation-of-constants formulation of the solution to Eq.~\ref{eq:trpmd};\cite{garcia1998long,hairer2000long,Sanz-Serna2008,McLachlan2014} 
specifically, the protocol in  Eq.~\ref{eq:Vextmolly} is a generalization of the mollified impulse method.\cite{garcia1998long}

Use of force mollification in the current work can  be motivated on physical grounds: In the absence of a physical potential, four of the considered integration schemes (OBABO, BAOAB, OBCBO, and BCOCB)  leave invariant the exact free ring-polymer Boltzmann-Gibbs distribution.\cite{KoBoMi2019} 
Therefore, the loss of any overlap between
the exact stationary distribution of the position marginals in the infinite-bead limit for  OBABO, BAOAB, and OBCBO must be attributed to the influence of the time evolution from the external potential in the schemes (i.e., the ``B'' sub-step) as implemented in Eq.~\ref{Bsubstep}; the BCOCB scheme does not suffer from this problem.
To remove this  pathology in the OBCBO scheme, we thus use mollification to taper down the external forces on the high-frequency modes, such that the resulting integration correctly reverts to free ring-polymer motion for those  modes, which should become decoupled from the external potential as the frequency increases. 
The specific appearance of the $1/2$ factor in the sinc function argument ensures that the sinc function switches from its high-frequency effect to its low-frequency effect when the period of the Matsubara frequency is commensurate with $\Delta t$; the zero-frequency ring-polymer centroid mode is untouched by mollification.

Force mollification requires only a small algorithmic modification of the OBCBO  integrator. 
Specifically, the ``B'' sub-step in Eq.~\ref{Bsubstep} is replaced with
\begin{equation}
\label{Msubstep}
\v{v} \leftarrow  \v{v} + \frac{\Delta t}{2} \frac{1}{m_n} \v{\tilde{F}}( \v{q} ),
\end{equation}
where the mollified forces are 
\begin{equation}
  \v{\tilde{F}}( \v{q} ) =   \sinc(\m{\tilde\Omega}\Delta t/2) \v{F}( \v{\tilde q} ) = \m{U}  \m{D}_{\Delta t} \m{U}^\textrm{T} \v{F}( \v{\tilde q} ) 
  \label{fmollified}
\end{equation}
where $\v{\tilde q} = \m{U} \m{D}_{\Delta t} \m{U}^\textrm{T} \v{q}$ are the mollified bead positions, and where $\m{D}_{\Delta t}$ is the diagonal matrix of eigenvalues associated with $\sinc(\m{\tilde\Omega}\Delta t/2)$, i.e., 
\begin{equation}
\label{diagomega}
\m{D}_{\Delta t} = \operatorname{diag}(\sinc(\tilde\omega_{0,n} \Delta t/2), \dots,  \sinc(\tilde\omega_{n-1,n} \Delta t/2) )
\end{equation}
where $\tilde \omega_{j,n}$ is the $j$th eigenvalue of $\m{\tilde\Omega}$.  In practice, the mollified forces are computed in normal mode coordinates as follows:
  \begin{enumerate}[(a)]
    \item Starting with the ring-polymer bead position in normal mode coordinates, obtain a copy of the mollified bead positions via 
        \begin{equation}\label{eq:qmol}
         \v{\tilde q} = \m{U} \m{D}_{\Delta t} \v{\varrho} \;.
     \end{equation}
    \item Evaluate the external forces at the mollified ring-polymer bead positions, $\v{F}(\v{\tilde q})$.
    \item Apply the remaining mollification to the forces in Eq.~\ref{fmollified} via 
       \begin{equation}
    \m{U}^\textrm{T} \v{\tilde{F}}( \v{q} ) = \m{D}_{\Delta t} \m{U}^\textrm{T} \v{F}(\v{\tilde q})  \;.
    \end{equation}   
\end{enumerate}
 \begin{comment}
 \begin{enumerate}[(a)]
    \item Starting with the ring-polymer bead positions in Cartesian coordinates, obtain a copy of the mollified bead positions via the series of matrix multiplies
        \begin{equation}\label{eq:qmol}
         \v{\tilde q} = \m{U} \m{D}_{\Delta t} \m{U}^\textrm{T} \v{q}
     \end{equation}
    \item Evaluate the external forces at the mollified ring-polymer bead positions, $\v{F}(\v{\tilde q})$
    \item Apply the remaining mollification to the forces in Eq.~\ref{fmollified} via the series of matrix multiplies
       \begin{equation}
    \v{\tilde{F}}( \v{q} ) = \m{U} \m{D}_{\Delta t} \m{U}^\textrm{T} \v{F}(\v{\tilde q})   \;.
    \end{equation}   
\end{enumerate}
\end{comment}
We emphasize that in comparison to the standard force update (Eq.~\ref{Bsubstep}) the use of the mollified force update (Eq.~\ref{Msubstep}) introduces neither additional evaluations of the external forces nor $n\times n$ matrix multiplies associated with the discrete Fourier transform; it  therefore avoids any significant additional computational cost.

This mollification scheme preserves reversibility and symplecticity as well as local-third order accuracy of the OBCBO scheme with timestep. We emphasize that the sinc-function-based mollification scheme in Eq.~\ref{Msubstep} is not unique, and alternatives can certainly be devised. 
Even within the  functional form of the mollification in  Eq.~\ref{Msubstep}, flexibility remains with regard to the choice of the matrix $\m{\tilde \Omega}$, which allows for mode-specificity in the way the mollification is applied.  A simple choice for this matrix is $\m{\tilde \Omega} = \m{\Omega}$, such that mollification is applied to all of the non-zero ring-polymer internal modes. 
With this choice, we arrive at a fully-specified integration scheme that replaces the original ``B'' sub-step in Eq.~\ref{Bsubstep} with the mollified force sub-step in Eq.~\ref{Msubstep}; we shall refer to this force-mollified version of OBCBO integration scheme as ``OMCMO." 
In the following sub-section, we propose a partially mollified choice for $\m{\tilde \Omega}$ that further improves the accuracy.

For the harmonic external potential, all of the previously derived relations for OBCBO (most notably Eqs.~\ref{eq:stab_1d}, \ref{eq:numerical_IM_1D_OBCOCBO_s}-\ref{eq:numerical_IM_1D_OBCOCBO_r}, and \ref{eq:permodeerror}-\ref{eq:permodevarerror}) also hold for OMCMO with $\Lambda$ suitably replaced by $\tilde \Lambda_j = \sinc^2(\omega_{j,n} \Delta t/2) \Lambda$.
Note that $\tilde \Lambda_j \le \Lambda$, since $\sinc^2(\mathsf{x}) \le 1$ for all $\mathsf{x} \ge 0$, making clear that the mollification reduces the effect of the external potential on the higher-frequency internal ring-polymer modes.

We now show that  mollifying the forces in the B substep fixes the pathologies of  OBCBO  in the infinite-bead limit, by restoring overlap between the sampled and exact  stationary distributions.  To see this, note that the $j$th summand in Eq.~\ref{eq:kakutani_series} for OMCMO satisfies \begin{align*}
\left( 1- \frac{s_{j}}{s_{j,\Delta t}} \right)^2  &\le \left( 1 - \frac{s_{j}^2}{s_{j,\Delta t}^2}  \right)^2  \le f( \omega_j \Delta t/2)   \frac{\Delta t^4 \Lambda^2 }{16 m^2} 
\end{align*}
where $f(\mathsf{x})=((1-\sinc^2(\mathsf{x}))/\mathsf{x}^2 + \sinc^2(\mathsf{x}))^2$, and we have used the infinite-bead limit for the ring-polymer internal-mode frequencies 
\begin{equation} \label{eq:continuous_eigenvalues}
\omega_{j} = \lim_{n \to \infty} \omega_{j,n} = \begin{cases}
\dfrac{\pi j}{\hbar \beta} 
 & \text{if $j$ is even} \;, \\
\dfrac{\pi (j+1)}{\hbar \beta}  & \text{else} \;.
\end{cases}
\end{equation} 
Since \[
\sum_{j=1}^{\infty} f(\omega_j \Delta t/2 )  \le    6 \frac{\hbar \beta}{\pi \Delta t} + 4 \;, 
\footnote{This inequality comes from using Eq.~\ref{eq:continuous_eigenvalues} to write 
$\sum_{j=1}^{\infty} f(\omega_j \Delta t/2 ) = \mathrm{I} + \mathrm{II}$ where $\mathrm{I}=2\sum_{j=1}^{\lfloor \hbar \beta / (\pi  \Delta t) \rfloor} f(j \pi \Delta t / (\hbar \beta) )$  and $\mathrm{II} = 2\sum_{j=\lceil \hbar \beta / (\pi \Delta t) \rceil}^{\infty} f(j \pi \Delta t / (\hbar \beta)  )
$. Then the first term admits the bound $\mathrm{I} \le 2 f(1) \hbar \beta/ (\pi \Delta t) < 4 \hbar \beta/ (\pi \Delta t)$, and for the second term we use 
$\mathrm{II} \le F(1) +\hbar \beta/(\pi \Delta t) \int_1^{\infty} F(x) dx$ where $F(\mathsf{x}) = 2 ((1-\sinc^2(\mathsf{x}))/\mathsf{x}^2 + 1/\mathsf{x}^2)^2$ is monotone decreasing on $[1, \infty)$ with $F(1) \le 4$ and $\int_1^{\infty} F(x) dx \le 2$.
}
\] we obtain \begin{equation}
\sum_{j=1}^{\infty} \left( 1-\frac{s_{j}}{s_{j,\Delta t}}  \right)^2 \le
\left( 6 \frac{\hbar \beta}{\pi \Delta t} + 4 \right) \frac{\Delta t^4 \Lambda^2 }{16 m^2} \;.
\end{equation}
Again invoking Kakutani's theorem (Eq.~\ref{eq:kakutani_series}),
it follows that the numerical stationary distribution has an overlap with the exact stationary distribution. As a byproduct of this analysis, we can also quantify the amount of overlap between the exact and numerically sampled stationary distributions,
\footnote{This quantification uses: (i) $d_{\mathrm{TV}} \le 2 d_{\mathrm{H}}$ where  $d_{\mathrm{TV}}$ is the total variation distance and
 $d_{\mathrm{H}}$ is the Hellinger distance; and (ii) subadditivity of the squared Hellinger distance, which implies that $d_{\mathrm{H}}^2(\mu, \mu_{\Delta t}) \le \sum_{j=1}^{\infty} d_{\mathrm{H}}^2(\mathcal{N}(0,s_j^2), \mathcal{N}(0,s_{j,\Delta t}^2)) \le \sum_{j=1}^{\infty} (1 - s_j^2 / s_{j,\Delta t}^2)^2 \le \left( 3 \hbar \beta / (\pi \Delta t) + 2 \right) \frac{\Delta t^4 \Lambda^2 }{8 m^2} $.}
 revealing that the total variation distance\cite{Gibbs2002} between these  distributions is given by  
 \begin{equation} \label{eq:tv_error}
    d_{\mathrm{TV}}(\mu, \mu_{\Delta t}) \le  \sqrt{\left( 6 \frac{\hbar \beta}{\pi \Delta t} + 4 \right)} \frac{\Delta t^2 \Lambda }{2 m}  \;.
 \end{equation} 
In summary, the force mollification strategy introduced here provably removes the pathologies due to the ``B'' sub-step in the case of a harmonic oscillator potential. 
Moreover, for any finite number of beads, the total variation distance between the exact and numerically sampled stationary distribution can be bounded by Eq.~\ref{eq:tv_error}, and thus, OMCMO admits error bounds that are dimension-free.

Before proceeding, we first return to Fig.~\ref{fig:logError} to compare the accuracy of  OMCMO with the un-mollified
OBCBO scheme for the internal-mode position marginal of the harmonic oscillator. 
As  seen  in Fig.~\ref{fig:logError}e for the results with a timestep of 1 fs,  the per-mode error obtained by the mollified scheme (OMCMO, green) decays more rapidly with mode number than does OBCBO.
Fig.~\ref{fig:logError}d further illustrates that upon summation of the per-mode contributions, the OMCMO prediction for the primitive kinetic energy  converges to a well-defined asymptote  with respect to the number of ring-polymer beads, whereas OBCBO diverges as discussed earlier. Similar behavior is seen for shorter MD timesteps (panels a-c), although the failure of OBCBO becomes less severe with this range of bead numbers as the timestep is reduced.

Although it is satisfying that mollification via  OMCMO  both formally and numerically ameliorates the problems of the OBCBO scheme 
in the high-bead-number limit, the OMCMO results in Fig.~\ref{fig:logError} are not ideal, since in some cases the OMCMO error is substantially larger than that of  OBCBO  when a modest number of beads is used (e.g., for 16 beads in panel d). This observation points to a simple and general refinement of the OMCMO scheme, which we discuss in the following subsection.

\subsection{Partial mollification}

Comparison of the per-mode errors from  OBCBO and OMCMO  in Fig.~\ref{fig:logError}e reveals that  lower errors for OMCMO are only enjoyed for internal modes that exceed a particular frequency (indicated by the vertical black line).  This observation suggests that if a ``crossover frequency" could be appropriately defined, then a refinement to OMCMO  could be introduced for which mollification is applied only to the ring-polymer internal modes with frequency that exceed this crossover value.

For the case of a harmonic external potential, this crossover frequency $\omega_{\mathrm{x}}$ can be found by comparing a bound for the per-mode error (Eq.~\ref{eq:permodevarerror}) for OBCBO 
\begin{equation} \label{j_error_OBCBO}
\rho_{j,\Delta t} \le \left( \frac{s^2_j}{m_n \omega^2_{j,n} \beta} \frac{\Delta t^2 \Lambda}{4 m - \Delta t^2 \Lambda} \right)
\end{equation} to that for OMCMO \begin{equation} \label{j_error_OMCMO}
\rho_{j,\Delta t} \le  g(\omega_{j,n} \Delta t /2 ) \left( \frac{s^2_j}{m_n \omega^2_{j,n} \beta} \frac{\Delta t^2 \Lambda}{4 m - \Delta t^2 \Lambda} \right),
\end{equation}
where $g(\mathsf{x})=(1-\sinc^2(\mathsf{x}))/\mathsf{x}^2 + \sinc^2(\mathsf{x})$.  Since $g(\mathsf{x}) \ge 1$ only when $\mathsf{x} \le 1$, we expect better accuracy if  mollification is only applied to those ring-polymer internal modes with frequencies $\omega_{j,n} \ge \omega_{\mathrm{x}}$, where $\omega_{\mathrm{x}} =  2/\Delta t$.  Although this result was derived for the case of a harmonic potential, it does not depend on $\Lambda$.   We call this resulting partly mollified integration scheme 
``OmCmO."   This scheme has the nice properties of OMCMO, including strong stability and dimensionality freedom.

Implementation of  OmCmO  is a trivial modification of OMCMO, requiring only that the diagonal elements of $\m{D}_{\Delta t}$ in Eq.~\ref{diagomega} are evaluated using
\begin{equation} 
\sinc(\tilde\omega_{j,n}\Delta t/2) = \begin{cases}
1 & \text{for}\ \omega_{j,n} < \omega_{\mathrm{x}} \\
\sinc(\omega_{j,n}\Delta t/2) & \text{otherwise},
\end{cases}
\end{equation}
where $j=0,\dots,n-1$.
In physical terms, the emergence of $2/\Delta t$ in the crossover frequency is intuitive, since as was previously mentioned, it corresponds to having the ring-polymer mode undergo a full period per timestep $\Delta t$.

Finally, numerical results for the case of a harmonic potential (Figs.~\ref{fig:logError}a-d) reveal that the partially modified OmCmO scheme (cyan) achieves both robust convergence of the primitive kinetic energy with increasing bead number, as well as better or comparable accuracy than the OBCBO and OMCMO integration schemes -- as expected.  However, it must be emphasized that for all panels of Fig.~\ref{fig:logError}, the BCOCB scheme (which requires no force mollification) is by far the most accurate and stable.

\section{Results for anharmonic oscillators}\label{sec:aHO_results}

Having numerically characterized the performance of the various non-preconditioned PIMD integrators for the case of the harmonic oscillator external potential in Fig.~\ref{fig:logError}, we now turn our attention to anharmonic external potentials.  In this section, we consider both a weakly anharmonic (aHO) potential
\begin{equation}
V(q)=\Lambda\left(\frac{1}{2}q^2+\frac{1}{10}q^3+\frac{1}{100}q^4\right)  \label{eq:anharm}
\end{equation}
and the more strongly anharmonic quartic potential
\begin{equation}
V(q)=\frac{1}{4}q^4.  \label{eq:quart}
\end{equation}
All calculations are performed using $\hbar=1$, $m=1$, and $\beta=1$.
Assuming the system to be at room temperature ($300$ K), then the thermal timescale corresponds to $\beta\hbar\approx25.5$ fs and 
$\Lambda=m\omega^2$, where  $\omega=3315$ cm$^{-1}$ 
for $\Lambda=256$.
The trajectories are performed with the centroid mode uncoupled from the thermostat (i.e., in the  manner of T-RPMD); for the remaining $n-1$ internal modes, simulations  performed with the OBABO and BAOAB schemes use the standard\cite{Ceriotti2010,Rossi2014} damping schedule of $\m{\Gamma}=\m{\Omega}$, 
and   simulations performed using the Cayley modification (i.e., BCOCB, OBCBO, OMCMO, and OmCmO) use  friction
$\gamma_j = \min(\omega_{j,n},0.9\gamma_j^{\textrm{max}}(\Lambda),0.9\gamma_j^{\textrm{max}}(0))$ for the $j^{\textrm{th}}$ mode, where $\gamma_j^{\textrm{max}}(\Lambda)$ is the friction that saturates the inequality in Eq.~\ref{eq:stab_1d}; for the  quartic potential, we set $\Lambda=1$  in this calculation of $\gamma_j^{\textrm{max}}$.

\begin{figure}
\begin{center}
\includegraphics[width=0.48\textwidth]{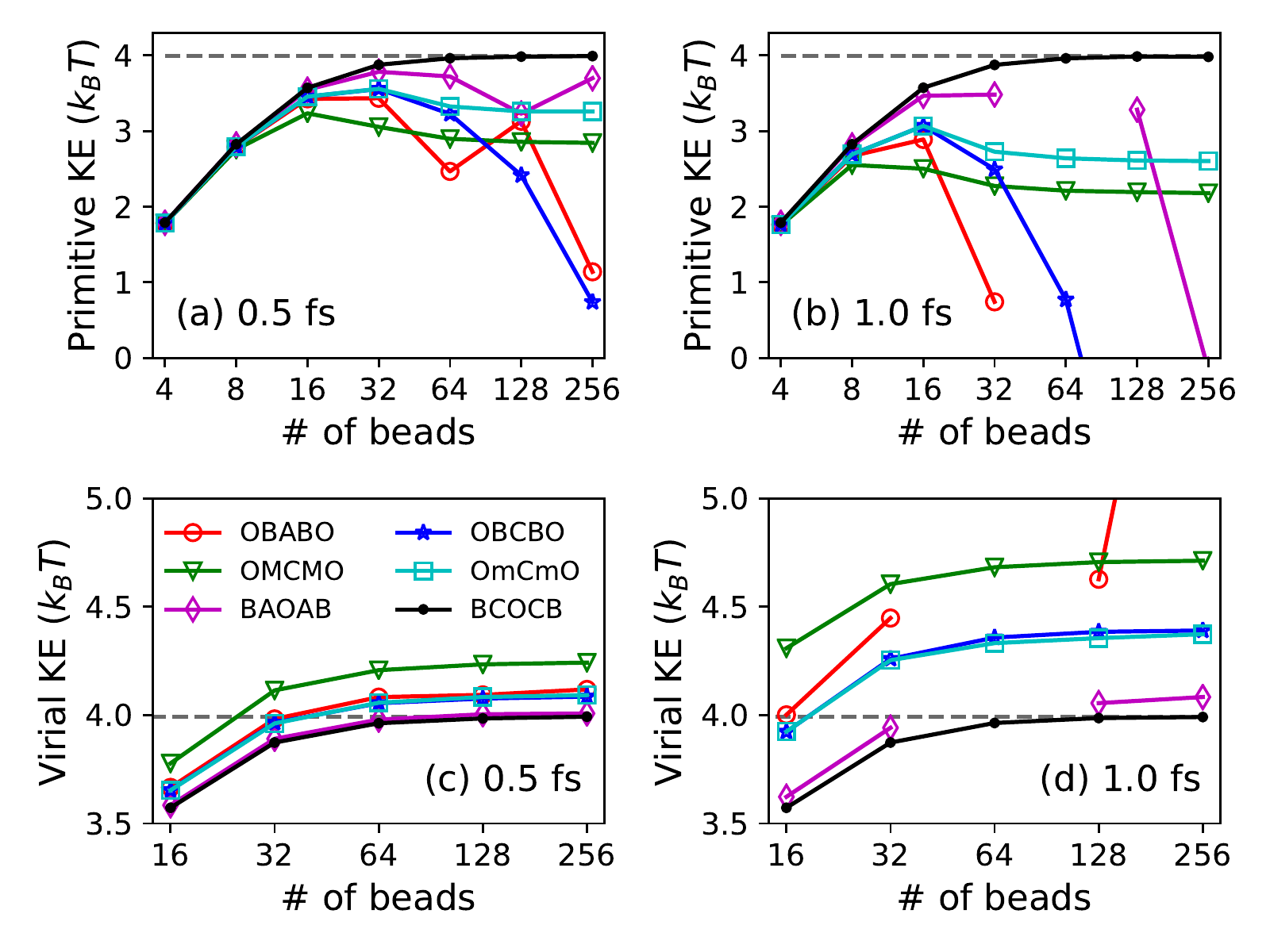}
\end{center}
\caption{\small  
{\bf Primitive and virial kinetic energy expectation values} as a function of bead number for the weakly anharmonic potential corresponding to $3315$ cm$^{-1}$ at room temperature, with results obtained using a timestep of 0.5 fs (a,c) and 1.0 fs (b,d). The standard error of all visible data points in each plot is smaller than the symbol size.
The exact kinetic energy is indicated with a dashed line. 
}
  \label{fig:Virial2}
\end{figure}

Figures \ref{fig:Virial2}a and b presents kinetic energy expectation values for the aHO potential corresponding to $3315$ cm$^{-1}$ at room temperature.  For the primitive kinetic energy expectation value, the results obtained using the various integration schemes with timesteps of both 0.5 fs (panel a) and 1.0 fs (panel b) are  consistent with the observations for the harmonic potential in Fig.~\ref{fig:logError}; specifically, the integrators without dimensionality freedom (OBABO, BAOAB, and OBCBO) fail to converge with increasing bead number, while the mollified integrators (OMCMO and OmCmO) smoothly converge with increasing bead number, and  the partially mollified scheme (OmCmO) is consistently more accurate than OBCBO and OMCMO.  However, it is also clear that  BCOCB  exhibits the best accuracy with increasing bead number, converging to the exact result without perceivable timestep error.

Figures \ref{fig:Virial2}c and d present the corresponding results for the virial kinetic energy expectation value,
\beq
\braket{KE_{\textrm{virial}}}=\frac{1}{2\beta} - \frac{1}{2}\left\langle(\v{q}-\bar{q})\cdot \v{F}(\v{q})\right\rangle
\label{eq:virialKE}
\eeq
where $\bar{q}$ is the centroid (bead-averaged) position. 
Whereas the virial kinetic energy for all of the strongly stable integration schemes is well behaved, the OBABO and BAOAB schemes perform erratically at large timesteps due to their provable non-ergodicities.\cite{KoBoMi2019}  Appealingly, the BCOCB scheme 
is consistently the most accurate for the virial kinetic energy expectation value, as it was for the primitive kinetic energy expectation value.  

\begin{figure}
\begin{center}
\includegraphics[width=0.46\textwidth]{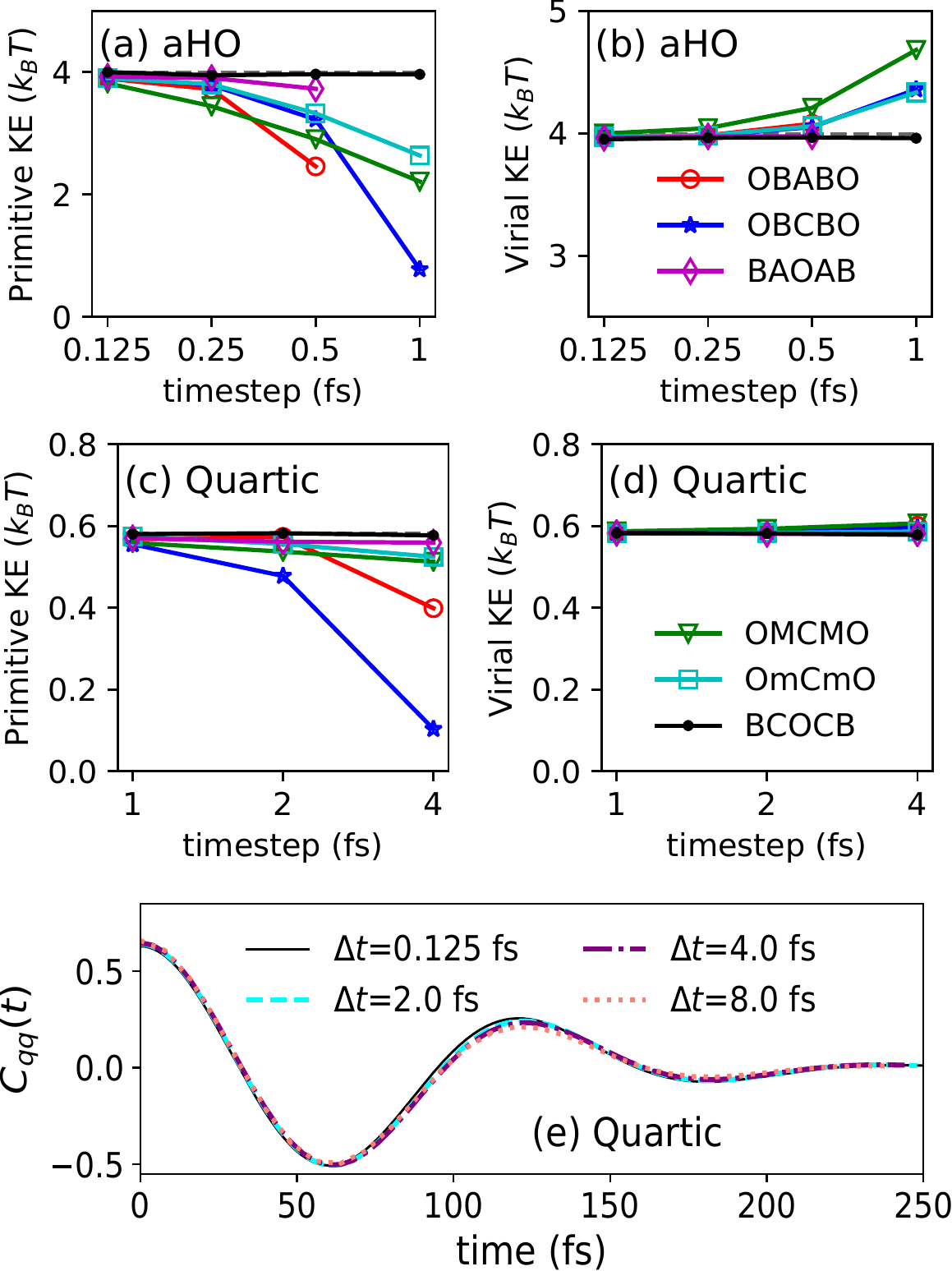}
\end{center}
\caption{\small  
{\bf Primitive and virial kinetic energy expectation values} as a function of the timestep for the weakly anharmonic potential corresponding to $3315$ cm$^{-1}$ at room temperature (a,b), 
and the quartic potential (c,d).  
The exact kinetic energy is indicated with a dashed line. The standard error of all visible data points in each plot is smaller than the symbol size.
Also, the position autocorrelation function (e) for the quartic oscillator at room temperature computed using T-RPMD with the BCOCB integrator. 
 Results are obtained using  64 ring-polymer beads using timesteps of $\Delta t= 0.125, 2, 4$, and $8$ fs.
}
  \label{fig:Time}
\end{figure}

Figure~\ref{fig:Time}a-d shows the results of the various numerical integration schemes for the primitive and virial kinetic energy expectation values, as a function of the MD timestep using 64 ring-polymer beads.  Results are shown for both the 
aHO and the strongly anharmonic quartic oscillator. 
In all cases, the BCOCB scheme is consistently the most accurate across this array of model systems.

Finally,  Fig.~\ref{fig:Time}e illustrates the use of the BCOCB integrator for the calculation of real-time quantum dynamics via T-RPMD, replacing the often-employed OBABO integration scheme. 
Using 64 beads, the T-RPMD results are plotted for a range of integration timesteps. 
Strikingly, over the entire range of considered timesteps,  BCOCB introduces negligible error in the calculated position time autocorrelation function; it is confirmed that these results are visually indistinguishable from those obtained using the OBABO integrator in the small-timestep limit.

\section{Results for liquid water}\label{sec:water_results}

The previous sections have demonstrated the strong performance of the BCOCB integrator  for obtaining both PIMD statistics as well as real-time dynamics via the T-RPMD model, in model systems.
Here, we test the accuracy and stability of the various un-mollified integration schemes (OBABO, OBCBO, BAOAB, and BCOCB) in liquid water, a high-dimensional and relatively complex system.  
Specifically, we consider a periodic $32$-molecule water box at a temperature of $298$ K and a density of $0.998$ g/cm$^3$, as described by the q-TIP4P/F force field.\cite{Habershon2009}

\begin{figure}
\begin{center}
\includegraphics[width=0.46\textwidth]{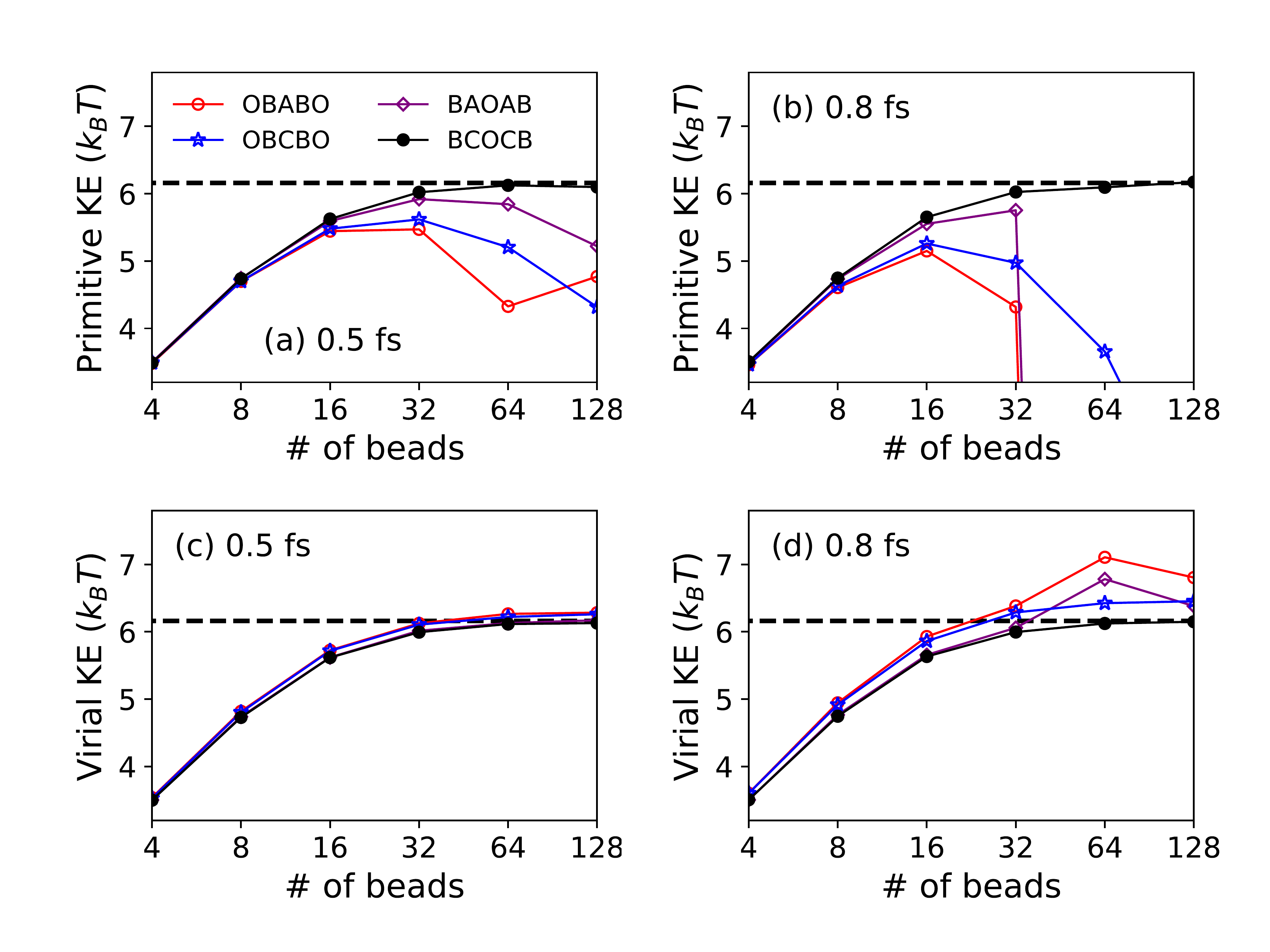}
\end{center}
\caption{\small  
{\bf Primitive and virial kinetic energy expectation values} as a function of the bead number per hydrogen atom in liquid water at 298 K and 0.998 g/cm$^3$ at timestep $\Delta t = 0.5$ fs (a, c) and $\Delta t = 0.8$ fs (b, d).
The reference kinetic energy, obtained from a converged staging PIMD simulation at timestep $\Delta t = 0.1$ fs and bead number $n = 256$, is indicated with a dashed line.
The standard error of all visible data points in each plot is smaller than the symbol size.
}
\label{fig:KE_water_numbeads}
\end{figure}

\begin{figure}
\begin{center}
\includegraphics[width=0.40\textwidth]{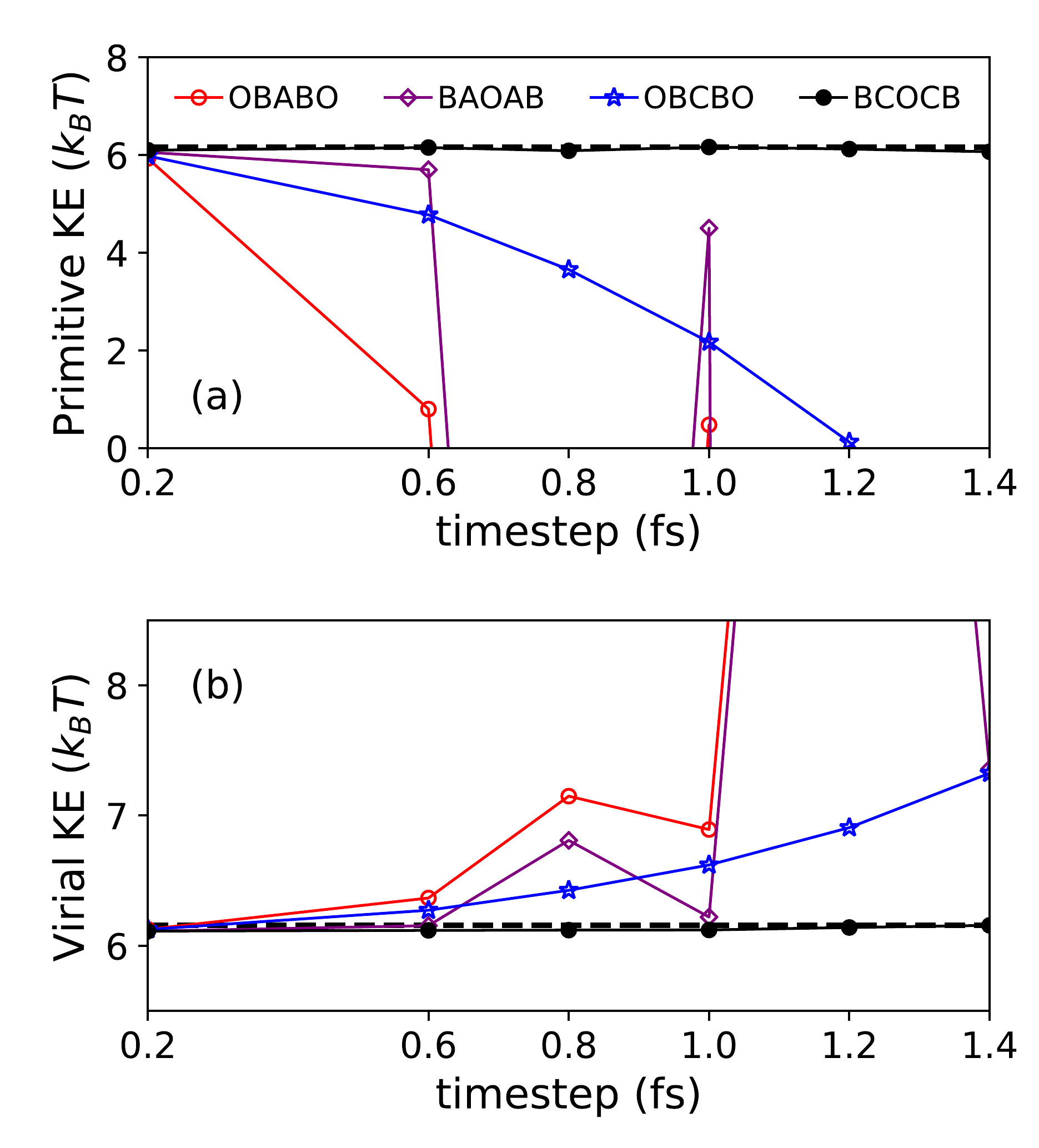}
\end{center}
\caption{\small  
{\bf Primitive and virial kinetic energy expectation values} as a function of the timestep per hydrogen atom in liquid water at 298 K and 0.998 g/cm$^3$, as described by a $64$-bead ring polymer.
The reference kinetic energy, obtained from a converged staging PIMD simulation at timestep $\Delta t = 0.1$ fs and bead number $n = 256$, is indicated with a dashed line.
The standard error of all visible data points in each plot is smaller than the symbol size.
\label{fig:KE_water_dt}
}
\end{figure}

In Fig.~4, we compare the accuracy achieved by the different integrators for the average kinetic energy per hydrogen atom as a function of the number of ring-polymer beads.
As in previous sections, we consider both the primitive (Eq.~\eqref{eq:primitive}) and virial (Eq.~\eqref{eq:virialKE}) estimators for the kinetic energy. 
For each choice of integrator, timestep, and bead number, the primitive and virial estimators for the kinetic energy of per hydrogen atom were averaged over a $1$-nanosecond trajectory integrated in the manner of T-RPMD, i.e., with the centroid mode uncoupled from the thermostat; for the remaining $n-1$ internal modes, simulations performed with the OBABO and BAOAB schemes use the standard\cite{Ceriotti2010,Rossi2014} damping schedule of $\bm{\Gamma} = \bm{\Omega}$, and simulations performed using the Cayley modification use friction $ \gamma_{j} = \min \{\omega_{j,n}, 0.9\gamma_{j}^\mathrm{max}(\omega_\mathrm{OH}^2), 0.9\gamma_{j}^\mathrm{max}(0)\}
$, where $\gamma_{j}^{\mathrm{max}}(\Lambda/m)$ saturates the inequality in Eq.~\eqref{eq:stab_1d} for the given values of $j$ and $\Lambda/m$ at the given time step, and  $\omega_\mathrm{OH}$ is the OH-stretch frequency from the harmonic bending force field term in the q-TIP4P/F force field.
Multi-nanosecond staging PIMD\cite{Tuckerman1993, Liu2016} simulations at a timestep of $0.1$ fs were performed to obtain a bead-converged reference value for the H-atom kinetic energy, plotted as a dashed line in Figs.~\ref{fig:KE_water_numbeads} and~\ref{fig:KE_water_dt}. 

The primitive kinetic energy expectation values in Figs.~\ref{fig:KE_water_numbeads}a and b show similar trends to those seen in Figs.~\ref{fig:logError} and~\ref{fig:Virial2} for the harmonic and weakly anharmonic oscillators.
For a $0.5$-fs timestep (Fig.~\ref{fig:KE_water_numbeads}a), at which all integrators exhibit strong stability for ring polymers with up to $64$ beads at the system temperature,\cite{KoBoMi2019} the OBABO, BAOAB, and OBCBO primitive kinetic energy estimates diverge from the converged result as the number of beads increases, in agreement with the proven result that the error in the ring-polymer configurational distribution generated with these schemes grows unboundedly with increasing bead number.
At the larger, $0.8$-fs timestep, (Fig.~\ref{fig:KE_water_numbeads}b), OBABO and BAOAB formally lose strong stability and their respective primitive kinetic energy estimates dramatically diverge for bead numbers greater than $32$; the strongly stable OBCBO scheme also yields a divergent result for the same reason as in Fig.~\ref{fig:KE_water_numbeads}a.
As seen on the HO and aHO model systems, the primitive kinetic energy expectation value from the BCOCB integrator monotonically converges to the reference
value  with increasing bead number, avoiding any perceptible timestep error.

Figs.~\ref{fig:KE_water_numbeads}c and d show the corresponding virial kinetic energy expectation values. For the smaller timestep of 0.5 fs, which is a common choice for path-integral simulations of water, all of the integrators perform similarly.  However, upon increasing the timestep to 0.8 fs,  significant differences in the performance of the integrators emerges, with only BCOCB avoiding perceptible timestep error.

To further compare the accuracy and stability of the OBABO, BAOAB, OBCBO, and BCOCB integrators, Fig.~\ref{fig:KE_water_dt} considers the average kinetic energy per hydrogen atom obtained using $64$ beads over a wide range of timesteps.
These results show that BCOCB remains remarkably accurate for timesteps as large as $1.4$ fs for liquid water, which corresponds to the limit of stability for Verlet integration of the centroid mode. 
In comparison, OBCBO diverges monotonically as the timestep increases, reaching unphysical values for the primitive expectation value and yielding sizable error ($20\%$) for the virial expectation value. 
The erratic performance of both OBABO and BAOAB is due to the emergence of numerical resonance instabilities at timesteps greater than $0.6$ fs at the employed bead number; indeed, the largest safe timestep at which OBABO and BAOAB remain strongly stable for $n = 64$, $\Delta t_\star \approx 0.63$ fs, precedes the range of timesteps in Fig.~\ref{fig:KE_water_dt} for which these integrators vary erratically.

\begin{figure}
\begin{center}
\includegraphics[width=0.40\textwidth]{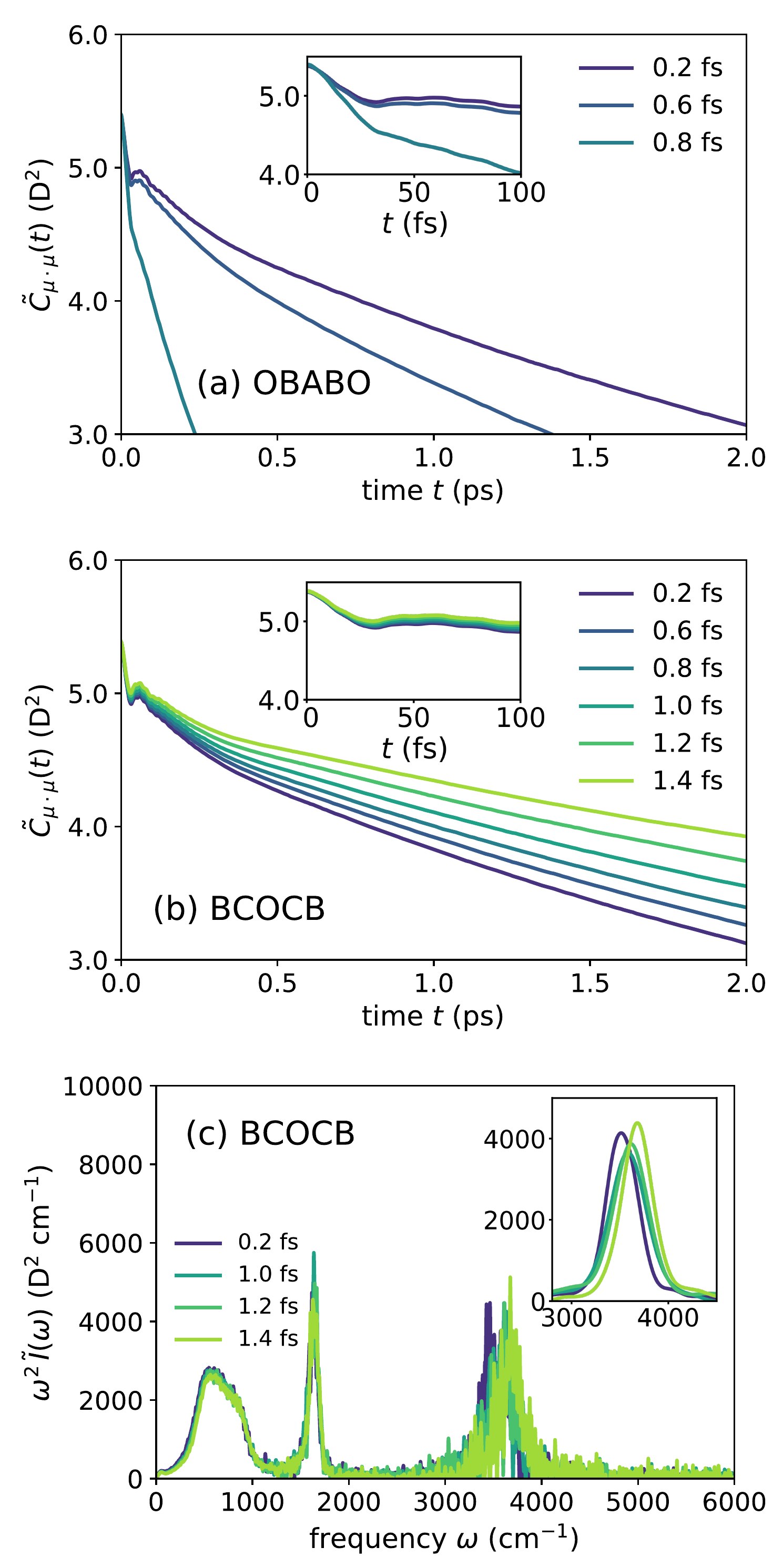}
\end{center}
\caption{\small  
{\bf Dynamical properties of liquid water} computed using T-RPMD with the (a) OBABO and (b,c) BCOCB integration schemes. 
Panels (a) and (b) present the Kubo-transformed dipole autocorrelation function computed with various timesteps, and panel (c) presents the absorption spectrum from the BCOCB correlation function at each timestep.  The inset to panel (c) presents the OH stretching region with smoothing.
\label{fig:waterdynamics}
}

\end{figure}

Extending beyond statistics, we now consider the dynamical properties of liquid water.
Given the superiority of the BCOCB scheme for the calculated statistical properties in Figs.~\ref{fig:KE_water_numbeads} and \ref{fig:KE_water_dt}, we present results that focus on this scheme in comparison to the most widely used OBABO scheme. 
In particular, we consider the liquid water infrared absorption spectrum,\cite{Habershon2008} which is proportional to $\omega^2 \tilde{I}(\omega)$ where the dipole spectrum $\tilde{I}(\omega) = \int_\mathbb{R} \mathrm{d}t \, e^{-i \omega t} \tilde{C}_{\mu\cdot\mu}(t)$ is the Fourier transform of the Kubo-transformed dipole autocorrelation function $\tilde{C}_{\mu\cdot\mu}(t)$.
The latter is approximated in the RPMD model by\cite{Miller2005} $\tilde{C}_{\mu\cdot\mu}(t) = \frac{1}{N} \sum_{i=1}^{N} \left\langle \bar{\mu}_i(t) \cdot \bar{\mu}_i(0) \right\rangle$, where $N$ is the number of molecules in the liquid, $\bar{\mu}_i(t)$ is the bead-averaged dipole moment of molecule $i$ at time $t$, 
and the angle brackets denote averaging over the ring-polymer thermal distribution.
To obtain the time-correlation functions and spectra shown in Fig.~\ref{fig:waterdynamics} for the OBABO and BCOCB integration schemes, $12$-nanosecond T-RPMD trajectories were simulated for a ring polymer with $64$ beads and timesteps ranging from $0.2$ to $1.4$ fs, using the same friction schedule as described for Figs.~\ref{fig:KE_water_numbeads} and \ref{fig:KE_water_dt}. 

Along each trajectory, the velocities of all degrees of freedom in the system were drawn anew from the Maxwell-Boltzmann distribution every $20$ picoseconds;
the autocorrelation function was evaluated out to $2$ picoseconds by averaging over staggered windows of that time-length within every $20$-picosecond trajectory segment; and exponential-decay extrapolation was used to extend the autocorrelation function before evaluating its numerical Fourier transform to obtain the infrared absorption spectrum.

Fig.~\ref{fig:waterdynamics}a and b present the dipole autocorrelation functions obtained using the OBABO and BCOCB integrators with a 
range of timesteps.
For the OBABO integrator, the calculated correlation function is qualitatively incorrect for timesteps as large as $0.8$ fs.
For the BCOCB integrator, the resulting correlations functions are far more robust with respect to timestep.
Although modest differences are seen in the exponential tail of the correlation function, the dynamics on vibrational timescales (see inset) is largely unchanged as the timestep is varied from $0.2$ fs to $1.4$ fs. 
Fig.~\ref{fig:waterdynamics}c further emphasizes this point by showing the absorption spectrum that is obtained from the BCOCB time-correlation functions with the various timesteps.
To minimize  bias, we avoided any smoothing of the spectra shown in panel c.
It is clearly seen that the librational and bending features (below $2500$~cm$^{-1}$) are visually indistinguishable over the entire range of considered timesteps.
To clarify the comparison for the stretching region above $3000$~cm$^{-1}$, we smooth the raw spectra in that region by convolution against a Gaussian kernel with a width of $150$~cm$^{-1}$ (see inset).
Again, the robustness of the simulated spectrum over this span of timesteps is excellent, with the only significant effect due to finite-timestep error being a slight blue-shifting of the OH stretching frequency for the results using a $1.4$-fs timestep, which is nearly three times larger than the typical value employed for the OBABO scheme for simulations with $64$ beads.
Taken together, these results indicate that the BCOCB integrator provides an excellent description of both PIMD statistics and T-RPMD dynamics in realistic molecular systems, substantially improving the accuracy and stability of previously employed numerical integrators.

%%%%%%%% SUMMARY %%%%%%%%%
\section{Summary}
\label{sum}

In a previous paper,\cite{KoBoMi2019} we showed that essentially all schemes for the non-preconditioned equations of motion of PIMD, including the widely used OBABO scheme, lack strong stability due to the use of exact free ring-polymer time evolution in the ``A'' sub-step, and we proved that this lack of strong stability gives rise to a lack of ergodicity in the thermostatted trajectories.  We further showed that ergodicity can be restored by simply replacing the ``A'' sub-step with the Cayley transform. 

In the current work, we show that a  completely distinct -- yet equally important -- pathology exists in the ``B" sub-step of previously developed non-preconditioned PIMD integrators, due to the outsized effect of the external potential on the dynamics of the high-frequency ring-polymer modes.
Specifically, we show that  previous integrators (including OBABO, BAOAB, and OBCBO) yield a numerical stationary distribution for which the overlap with the exact stationary distribution vanishes in the infinite-bead limit.
We then show that this pathology is completely avoided in the BCOCB scheme, and we further show that the pathology can be eliminated for the OBCBO scheme by  
suitably mollifying the ``B'' sub-step, yielding the dimension-free non-preconditioned PIMD integrators, namely BCOCB, OMCMO, and OmCmO. 
Implementation of the dimension-free  integration schemes involves no significant additional computational cost, 
no additional parameters, and no increase in algorithmic complexity in comparison to either OBABO or BAOAB.
Furthermore, since the integrators considered here are all non-preconditioned, they can immediately be used for computing the equilibrium statistical properties as well as dynamical properties via the RPMD model. 
The numerical performance of the BCOCB scheme is particularly striking, yielding results that are markedly better in terms of accuracy and timestep stability than any of the other considered integrators. For liquid water, it is shown that BCOCB allows for timesteps as large as 1.4 fs while exhibiting  minimal timestep error in the calculation of both equilibrium expectation values and the  dipole absorption spectrum.

\begin{acknowledgements}
We thank Andreas Eberle and Ondrej Marsalek for helpful discussions.
N.~B.-R.~acknowledges support from the National Science Foundation under Award No.~DMS-1816378 and the Alexander von Humboldt foundation.
R.~K., J.~L.~R.-R.~and T.~F.~M.~acknowledge support from the Department of Energy under Award No.~DE-FOA-0001912 and the Office of Naval Research under Award No.~N00014-10-1-0884.
\end{acknowledgements}

\section{Appendix A: Other splittings}
\label{app:other_splittings}

There are exactly four locally third-order accurate symmetric splitting schemes that involve one new force evaluation per integration step, and that involve splitting the T-RPMD dynamics into Hamiltonian and thermostat parts: OBABO, BAOAB, OABAO,and ABOBA.  In Section \ref{sec:BCOCB}, we quantified the  properties of OBABO, BAOAB and their Cayley-modifications in the case of a harmonic external potential. The corresponding properties of the Cayley modifications of OABAO and ABOBA are given below.  
\begin{itemize}
    \item OCBCO is exact in the velocity marginal, but the variance in the position marginal is $(\beta m_n)^{-1} s_{j, \Delta t}^2$ where $s_{j, \Delta t}^2 = (4 m - \Delta t^2 \Lambda) / (4 \Lambda + 4 m \omega_{j,n}^2)$;
    \item CBOBC is exact in the position marginal, but the variance in the velocity marginal is $(\beta m_n)^{-1} r_{j, \Delta t}^2$ where $r_{j, \Delta t}^2 = 4 m / ( 4 m - \Delta t^2 \Lambda)$.
\end{itemize}
Numerical experiments confirmed these properties but did not show significant improvement in accuracy compared with BCOCB.   Therefore, we did not include numerical results for these schemes.  

\bibliography{Mollify,RPMD}

\end{document}